\documentclass[referee]{raa}            

\usepackage{graphicx,times}             
\usepackage{natbib}
\usepackage{amssymb,amsmath}
\bibpunct{(}{)}{;}{a}{}{,}

\begin{document}

   \title{Suggested quasi Cassegrain system for multi-beam observation of FAST
}

   \volnopage{Vol.0 (20xx) No.0, 000--000}      
   \setcounter{page}{1}          

   \author{Ding-qiang Su
      \inst{1,2}
   \and Hua Bai
      \inst{2}
   \and Xiangqun Cui
      \inst{2}
   }

   \institute{School of Astronomy and Space Science \& Key Laboratory of the Ministry of Education,
Nanjing University, Nanjing 210023 , China; {\it dqsu@nju.edu.cn}\\
        \and
             National Astronomical Observatories/Nanjing Institute of Astronomical Optics and Technology (NIAOT),
Chinese Academy of Science, 188 Bancang Street, Nanjing 210042, China\\
\vs\no
   {\small Received~~20xx month day; accepted~~20xx~~month day}}

\abstract{ FAST, the largest single-dish radio telescope in the world, has a 500-meter diameter main reflector and a 300-meter diameter illumination area. It has a shape variable main reflector, which changes the shape of the illuminated area in the main reflector into a paraboloid continuously . In this article, we propose a quasi Cassegrain system to FAST. The detailed design results are given in this paper. Such a quasi Cassegrain system only needs to add a 14.6-meter diameter secondary reflector, which is close to the size of the feed cabin, the distance from the secondary reflector to the focus is only 5.08-meter, and it has excellent image quality. In this quasi Cassegrain system the shape of the illuminated area in the main reflector continuously changes into an optimized hyperboloid. Using this quasi Cassegrain system from frequency 0.5 G to 8 G, the multi-beam system can include 7 to 217 feeds. If this system is used in combination with PAF technology, more multi-beam feeds can be used.
}

   \authorrunning{Ding-qiang Su, Hua Bai \& Xiangqun Cui }            
   \titlerunning{Suggested quasi Cassegrain system for FAST}  

   \maketitle

%
%
\section{Introduction}           
\label{sect:intro}

FAST- Five-hundred-meter Aperture Spherical Radio Telescope \citep{FASTweb, Peng09, Nan17, Jiang19} is the Chinese national large scientific engineering project and the largest single-dish radio telescope in the world. Active optics \citep{Wilson99,Lemaitre09}, which was developed in the 1980s and is used for overcoming gravitational deformation and thermal deformation in large telescopes, keeps the large mirrors of telescopes in accurate shape and collimation. In 1986 Ding-qiang Su put forward the "shape variable mirrors" idea \citep{Su86}, using active optics technology to achieve some optical systems which could not be realized traditionally. This opened up a new direction for active optics. In that paper one of the two examples is as follows:

\emph{"Like Arecibo 305-meter radio telescope, a very large optical telescope, as proposed by a few people, could use a fixed or half fixed spherical primary mirror and let the secondary or corrector do tracking movement near the primary focus plane. In these configurations the continuous fields of view are very small due to the considerable off-axis aberrations. If we use the method described in this paper, i.e., continuously changing the shape in the area in use on the primary to make it maintain the optimum shape, e.g. paraboloid or hyperboloid, consequently the best correction could be obtained in large continuous field." }

Although the above idea was proposed for optical systems, since optical systems and radio systems tend to borrow from each other, the above idea also can be used for radio systems. In 1998 Yuhai Qiu \citep{Qiu98AA, Qiu98MN}, one of the main members of FAST, proposed to continuously change the shape of the illuminated area in the 500 m main reflector into a paraboloid, i.e., the same idea as that of our paper \citep{Su86}. Bo Peng, main member of FAST, actively supported this innovative idea. Finally, this idea was adopted for FAST. This is the most difference between FAST and Arecibo and the key innovation in FAST. The "shape variable mirrors" idea also is the key innovation in LAMOST, another Chinese national large scientific engineering project and the largest wide field of view combining large aperture optical telescope in the world \citep{LAMOSTweb,Wang96,Cui12}.
In this article we propose to add a quasi Cassegrain system for FAST. The detailed process of design and results are given in the following.


\section{A brief introduction to FAST}
\label{sect:FAST}
\begin{figure}
\centering
\includegraphics[width=\textwidth, angle=0]{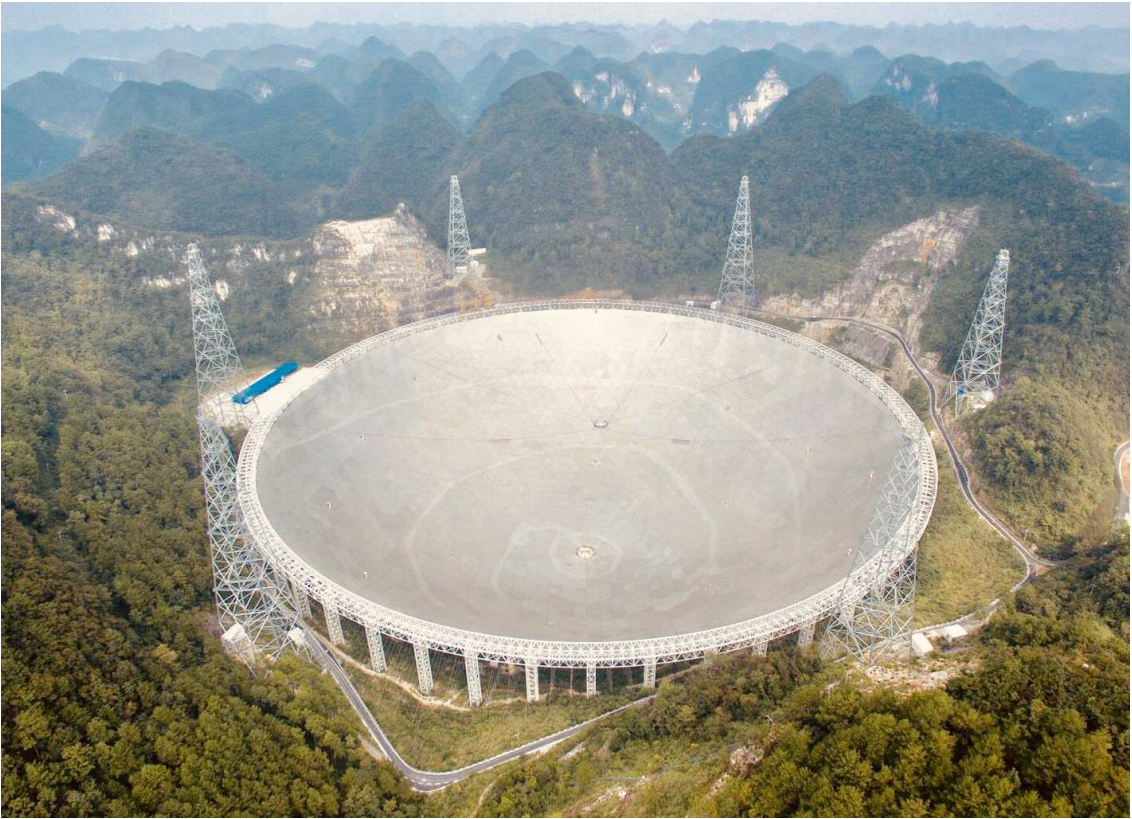}
\caption{The photo of FAST provided by FAST team}
\label{fig:FAST}
\end{figure}

In 1993, an international astronomical community (10 countries including China) proposed to develop a next-generation Large Telescope (LT) for radio astronomy. As an LT scheme, in 1994 some Chinese astronomers proposed to develop a radio telescope similar to Arecibo \citep{Areciboweb}, with a spherical main reflector, a 500m aperture, and a 300m diameter illuminated area. Since the shape of Arecibo's main reflector is spherical, it has serious spherical aberration. If addition of a corrector to eliminate the spherical aberration, the field of view (FOV) is still very small due to considerable off-axis aberrations. If the f-ratio (the focal length divided by diameter of illuminated area in main reflector) is increased these aberrations can be decreased, but the observable sky range will lessen as well. In Arecibo, the diameter of the main reflector is 305m, the diameter of the illuminated area is 221m, the radius of curvature of the main reflector is 265m, the f-ratio is 0.600. See Section 1 of this paper for the "shape variable mirrors" idea \citep{Su86}, i.e., continuously changing the shape in the illuminated area on the main reflector to make it maintain a paraboloid, so that the spherical aberration is eliminated completely.
Since the illuminated area in main reflector of FAST is paraboloid, it has no spherical aberration, therefore it can do observation without any corrector. In FAST the f-ratio is 0.4611. Thus, FAST can observe a much larger sky range than Arecibo, which is another important advantage. The first light of FAST was on September 25, 2016 and now it is in commissioning stage. Australian astronomers are helping the FAST team to develop multi-beam feeds. A 19-beam feed, developed by Australia, has been installed and has started to work at the FAST paraboloid prime focus. Until June 10, 2019, 100 new candidates of pulsars have been found using drift scanning method, and among them 65 have been identified as pulsar. A pulsar with interesting emission properties have been discovered \citep{Zhang19}. Figure~\ref{fig:FAST} is a photo of FAST.

\section{The quasi Cassegrain system for FAST}
\label{sect:CASS}

\begin{figure}
\includegraphics[width=0.75\textwidth, angle=0]{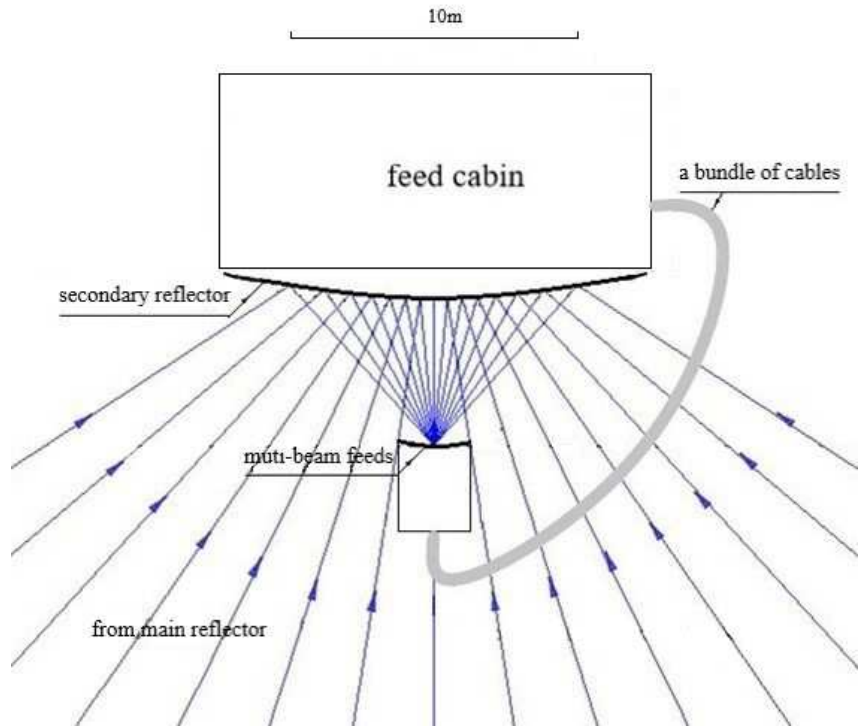}
\centering
\caption{FAST quasi Cassegrain system, secondary reflector and multi-beam feeds }
\label{fig:CASSFAST}
\end{figure}

In paper \citep{Su86} writes that the shape of the illuminated area on the main reflector also can be hyperboloid. This means a secondary reflector is added and can obtain a wide FOV quasi Cassegrain system for multi-beam observation. Hua Bai has designed a series of quasi Cassegrain systems for FAST, of which secondary reflectors have different magnifications and all of these diameters are near 15m, i.e., near the diameter of the feed cabin. During the design process we were clearly aware that this quasi Cassegrain system is a reflecting system, to which wavelength is irrelevant, and for the FOV a radius of 0.56 degree is taken. The coefficients $a_{4}, a_{6}, a_{8}, a_{10}$ of the secondary reflector shape formula in Table~\ref{tab:T1}, and the conic constant cc ($= -e^{2}$) of the illuminated area in the main reflector are optimized. The ZEMAX software is used for this design. Bai found magnification of the secondary reflector from 1.5-2 is good. As magnification increases the image quality improves, and the linear obstruction ratio is somewhat reduced, but the distance from the vertex of secondary reflector to focus is increased. The structure of the quasi Cassegrain system, which we selected, is shown in Table~\ref{tab:T1} and Figure~\ref{fig:CASSFAST}. The magnification of the secondary reflector is 1.628 and the quasi Cassegrain system's f-ratio is 0.7507. The shape of the illuminated area is hyperboloid with the conic constant cc (i.e. $= -e^{2}$) = -1.0311329, e is eccentricity. The maximum deviation between this hyperboloid and paraboloid is only 1.8cm in the illuminated area in the main reflector. Since the main reflector is a "shape variable mirror" the exchange between two shapes is easy. The diameter of the secondary reflector is 14.60m, which is close to the size of the feed cabin. This secondary reflector can be fixed under the feed cabin. In this quasi Cassegrain system the vertex of the secondary reflector to focus is 5.080 m. The multi-beam feeds, they are at the focus, can be connected to secondary reflector by leaf blade. The distance between the prime focus (i.e., the paraboloid focus) and the vertex of the secondary reflector is 3.03m. The RMS wave aberration with frequency G and FOV radius \emph{w} are given in Table~\ref{tab:T2}. The unit of the FOV radius \emph{w} is the same as the diameter of the Airy disk for corresponding frequency. Thus, it is easy to see how many feeds can be included for this multi-beam system. For example, in Table~\ref{tab:T2} if \emph{w}=3, frequency = 1G (i.e.$\lambda= 30cm$), RMS wave aberration is 0.0432$\lambda$. \emph{w}=3 means that the FOV radius is 3 times diameters of the Airy disk of frequency 1G. If the feed diameter is taken as the diameter of the Airy disk, 37 multi-beam feeds can be installed. We restrict that the
diameter of multi-beam feeds not exceeding 4m. The equivalent
the linear obstruction ratio (denotes as $\eta$) is 0.22. In this situation the energy loss is only about 5$\%$, and the diffraction distribution has a little deterioration. We stipulate that the image quality RMS wave aberration be less than $\lambda$/14 (i.e. $< 0.0714\lambda$) and the linear obstruction ratio be less than 0.22(i.e.$\eta<0.22$). In Table~\ref{tab:T2}, all listed values their the linear obstruction ratio $\eta<0.22$.  From Table~\ref{tab:T2} one can find that for frequency 0.5G a multi-beam system can include 7 feeds, $\eta=0.18$; for 1G, 37 feeds, $\eta=0.21$; for 2G, 91 feeds, $\eta=0.17$; for 4G, 169 feeds, $\eta=0.11$ and for 8G, 217 feeds, $\eta=0.064$, all these multi-beam systems satisfy the above two conditions. This quasi Cassegrain system has excellent image quality, many feeds can be included for multi-beam observation, and it is small and exquisite. We don't think this quasi Cassegrain system with secondary reflector is complicated. Since the main reflector of Arecibo is invariably spherical, when the secondary and tertiary reflectors are added, if its illuminated area is the same as ours the aperture of second reflector will be nearly three times bigger than ours. At the paraboloid prime focus coma is serious. The size of coma is proportional to the FOV. The diameter of the Airy disk is proportional to $\lambda$. From these one can find that for the paraboloid prime focus for a certain \emph{w} the RMS wave aberration has the same coefficient, which is irrelevant to $\lambda$. This is also demonstrated by calculation. But the paraboloid prime focus also has other off-axis aberrations (for example, astigmatism, high-order aberrations etc.), which are not proportional to FOV, so the above results are approximative. The RMS wave aberration of the FAST paraboloid system is also shown in Table~\ref{tab:T2}.
 \begin{table}
 \scriptsize
\bc
\begin{minipage}[]{100mm}

\caption[]{The structure of the FAST quasi Cassegrain system \label{tab:T1}}\end{minipage}
\setlength{\tabcolsep}{2.5pt}
\begin{tabular}{lcrcccccc}
 \hline\noalign{\smallskip}
Element & Aperture & Vertex radius  & Thickness & Conic constant &$a_{4}$ &$a_{6}$  & $a_{8}$ & $a_{10}$  \\
 & & of &   & cc & & & & \\
  &(m)& curvature(m) &(m) &(= -$e^{2}$)& & & & \\
 \hline\noalign{\smallskip}
Main reflector& 300 & -276.6600 & -135.3000	& -1.0311329& & & & \\
Secondary reflector & 14.60$^{\ast}$ & -15.70926 & 5.080000 & 0 & 8.884461E-13 &-2.095897E-20 &3.004833E-28 & -1.707826E-36\\
Focal surface & 4.00 &-5.518697 & &   & &  &  & \\
\noalign{\smallskip}\hline
\end{tabular}
\ec
\tablecomments{0.9\textwidth}{$^{\ast}$ corresponding to the diameter of linear FOV 4m.\\
$x=(c(y^{2}+z^{2}))/(1+\sqrt{(1-c^{2} (1+cc)(y^{2}+z^{2} )} ))+a_{4} (y^{2}+z^{2} )^{2}+a_{6} (y^{2}+z^{2} )^{3}+a_{8} (y^{2}+z^{2} )^{4}+a_{10} (y^{2}+z^{2} )^{5}$}
\end{table}

\begin{table}
\scriptsize
\bc
\begin{minipage}[]{100mm}
\caption[]{The RMS wave aberration with frequency and \emph{w} in FAST\label{tab:T2}}\end{minipage}
\setlength{\tabcolsep}{2.5pt}
\begin{tabular}{lcrcccccccccccccc}
 \hline\noalign{\smallskip}
Element&Frequency& & & & &RMS&wave&aberration& & & & & & \\
 & &\emph{w}=0&\emph{w}=1&\emph{w}=2&\emph{w}=3&\emph{w}=4&\emph{w}=5&\emph{w}=6&\emph{w}=7&\emph{w}=8&\emph{w}=9&\emph{w}=10&\emph{w}=11&\emph{w}=12\\
 \hline
 Paraboloid&Any& & & & & & & & & & & & & \\
 prime system&Frequency&0.0033&0.1031&0.2059&0.3088&0.4116&0.5145&0.6174&0.7202&0.8231&0.9259&1.029&1.132&1.234\\
 \hline
    &0.5G& & & & & & & & & & & & & \\
    &($\lambda$60cm)&0.0031&0.0098& & & & & & & & & & & \\
    & 1G& & & & & & & & & & & & & \\
    &($\lambda$30cm)&0.0061&0.0089&0.0197&0.0432& & & & & & & & & \\
  Cassegrain&2G& & & & & & & & & & & & & \\
    &($\lambda$15cm)&0.0123&0.0136&0.0178&0.0260&0.0393&0.0591&0.0864& & & & & & \\
   &4G& & & & & & & & & & & & & \\
    system&($\lambda$7.5cm)&0.0245&0.0252&0.0271&0.0306&0.0357&0.0427&0.0519&0.0638&0.0785&0.0967&0.1182&0.1436& \\
     &8G& & & & & & & & & & & & & \\
    &($\lambda$3.75cm)&0.0490&0.0493&0.0503&0.0520&0.0543&0.0573&0.0612&0.0657&0.0712&0.0778&0.0853&0.0940&0.1039\\
\hline\noalign{\smallskip}
\end{tabular}
\ec
\tablecomments{0.9\textwidth}{RMS wave aberration unit is $\lambda$. Reference value $\lambda$/14= 0.0714$\lambda$\\
\emph{w} is FOV radius. Unit of \emph{w} is Airy disk diameter of corresponding frequency.}
\end{table}
Phased Array Feed (PAF) is a very important new technology. In the long run, with this technology there will be no requirement for image quality for multi-beam systems, but at the moment, with PAF technology the wave aberration can be greater than $\lambda$/14 but not too much more. From Table~\ref{tab:T2} one can see that for the same \emph{w}, the RMS wave aberration of the quasi Cassegrain system is much smaller than that of the paraboloid prime focus system, which gives us the idea that by using the quasi Cassegrain system and PAF technology a multi-beam system can be obtained which includes more feeds, especially for high frequency. For FAST 0.5G and 1G, if more than 7 and 37 feeds are needed respectively, such multi-beam systems should be installed at the prime focus with PAF technology. For the above quasi Cassegrain system, if using PAF technology for frequency 2G, 127 feeds, $\eta=0.20$, for 4G, 271 feeds $\eta=0.17$, 8G, 469 feeds, $\eta=0.094$ may be obtained, and conceivably for 4G especially for 8G more feeds can be included by using PAF technology.

\section{THE FEASIBILITY OF QUASI CASSEGRAIN SYSTEM FOR FAST}
\label{sect:FEASIBILITY}
a. The shape of secondary reflector is invariant. The diameter of secondary reflector is only 14.6 m, and the distance to focus is 5.08 m.

b. The minimum working wavelength is about 3.75 cm(8G), so the surface tolerances of the secondary reflector is about RMS 0.5 mm, therefore the secondary reflector is easy to manufacture.

c. The surface shape of the secondary reflector is continuous and smooth.

d. The maximum deviation between the hyperboloid and the paraboloid on the illuminated area of the main reflector is only 1.8 cm, so it is easy to exchange these two shapes.

e. According to our preliminary structure analysis, the weight of secondary reflector is not more than 10 tons. If our suggested quasi Cassegrain system is adopted, the detailed engineering structure analysis will be implemented.

f. When the secondary reflector decenters by 5 cm, the mean value of the RMS wave aberration in FOV only increases $8\%$ for any wavelength. When the secondary reflector tilts in $2^\circ$, the mean value of the RMS wave aberration in FOV only increases $7\%$ for any wavelength. So, for given image quality, the tolerances of the secondary reflector are very loose, except that the image position movement caused by decenter and tilt of the secondary reflector needs to be corrected by auto-guiding system.

According to above analysis, our suggested quasi Cassegrain system for FAST is feasible and with low cost obviously.
\section{DISCUSSION}
a. Some radio telescopes also have a Cassegrain or quasi Cassegrain system. In general, their focus is fixed and near the vertex of main reflector, but there is no such a fixed focal position near the main reflector in FAST. So, for our quasi Cassegrain system, the focus must be close to the secondary reflector and with fixed position combined. In this case, according to our calculation, the image quality is still excellent.

b. One can notice that this quasi Cassegrain system increases the linear obstruction. However, when magnification is increased, the linear obstruction ratio can be reduced somewhat; when the size of the secondary reflector is increased without changing its magnification, the linear obstruction ratio can be also reduced.

c. The above quasi Cassegrain system also could serve as a reference for fully steerable radio telescopes with an aperture of several dozen meters to about one hundred meters.
\section{Conclusion}
Our suggested quasi Cassegrain system for FAST is feasible. It is useful for multi-beam observation. It is also useful if combining PAF technology with this quasi Cassegrain system for multi-beam observation.

\begin{acknowledgements}
We thank Professors Ron Ekers, Lister Staveley-Smith, Fu Han, Xingwu Zheng and Jinlin Han for their valuable discussion, and Professor Bo Peng's support. We also thank Dr. Weixing Su for improving the English for our paper.
\end{acknowledgements}

\label{lastpage}

\end{document}